\documentclass{svjour3}
\usepackage{fix-cm}
\usepackage{amsmath}
\def\be{\begin{equation}}
\def\ee{\end{equation}}
\def\bq{\begin{eqnarray}}
\def\eq{\end{eqnarray}}

\journalname{General Relativity and Gravitation}

\begin{document}

\title{Radiating stars with exponential Lie symmetries}

\author{R. Mohanlal \and S. D. Maharaj  \and  Ajey K. Tiwari \and R. Narain}

\institute{
R. Mohanlal \and S. D. Maharaj  \and  A. K. Tiwari \and R. Narain \at
Astrophysics and Cosmology Research Unit,
 School of Mathematics, Statistics and Computer  Science,
 University of KwaZulu-Natal, Private Bag X54001, Durban 4000, South Africa\\ \\
\\ \\
\email{maharaj@ukzn.ac.za} \\ \\
}
\date{Received: date / Accepted: date}

\maketitle

\begin{abstract}
We analyze the general model of a radiating star in general relativity. A group analysis of the under determined, nonlinear partial differential equation governing the model's gravitational potentials is performed. This analysis is an extension of previous group analyses carried out and produces new group invariant solutions. We find that the gravitational potentials depend on exponential functions owing to the choice of the Lie symmetry generator. The fundamental boundary equation to be solved is in general a Riccati equation. Several new exact families of solutions to the boundary condition are generated. Earlier models of Euclidean stars and generalized Euclidean stellar models are regained as special cases. Linear equations of state can be found for shear-free and shearing spacetimes.


\end{abstract}

\section{Introduction}
Initial models of collapsing radiating stars were highly idealised. An early exact model of a radiating stellar object was found by Kolassis et al. \cite{1} which regains the Friedmann dust model in the absence of heat flow. For a more realistic model it is important to include dissipative effects due to heat flow, shear viscosity, bulk viscosity, the electromagnetic field and a superposition of different relativistic fluids. The effects of dissipation have been highlighted in the recent investigations of Sharma and Das \cite{2}, Sarwe and Tikekar \cite{3} and Reddy et al. \cite{4}. Dissipation has a substantial effect on gravitational collapse for both the streaming out and the diffusion approximations. In addition, the role of all the kinematical quantities, namely the acceleration, the expansion and the shear, cannot be neglected. The treatments of Thirukkanesh et al. \cite{5}, Govender et al. \cite{6}, Govender et al. \cite{7} and Herrera et al. \cite{8} show how the dynamical evolution and the physical features depend on the kinematical and physical quantities.

The presence of acceleration, expansion and shear is the general case but these quantities increase the nonlinearity of the resulting system of partial differential equations. Chan \cite{9,10,11} and Pinheiro and Chan \cite{12,13} investigated dissipative effects, luminosity, particle production and temperature profiles using numerical methods. Thirukkanesh et al. \cite{5} found exact solutions to the boundary condition with all three kinematical quantities present. Euclidean stars in general relativity, fall in this category of spacetimes, and were first studied by Herrera and Santos \cite{14}. An exact model for an Euclidean star was presented by Govinder and Govender \cite{15}. A generalized class of Euclidean stars, containing metrics with an equation of state, was discussed by Abebe et al. \cite{16} using the Lie method of infinitesimal generators. In a recent approach Ivanov \cite{17} introduced a horizon function that transforms the boundary condition to a simpler form; exact solutions also arise in this approach.

Symmetry analysis is  a powerful method for solving  nonlinear differential equations by using the invariance properties of equations under Lie point transformations. Many physical models, inclusive of the current model, are governed by highly nonlinear differential equations that cannot be solved completely by simple techniques. Symmetry analysis has found much favour in solving these complicated models. In the setting of theoretical physics, the Lie symmetry approach has been successful in producing  physically viable models, ranging from quantum mechanical systems \cite{quant} to general relativistic systems \cite{msomi}. Solutions found via symmetry transformations often contain known solutions as special cases. 

An interesting feature of our analysis is that exponential functions arise in the symmetry analysis which was not the case in previous analyses of radiating stars invariant under Lie symmetries.

The primary objective of this investigation is to use Lie symmetry analysis to find new exact solutions to the boundary condition of  a radiating star. The structure of this paper is as follows: In Sect. \ref{sect2}. we present the metrics describing the interior and exterior of a radiating sphere. We discuss the conditions that the metrics must satisfy on the boundary. We show that the boundary condition can be written as a nonlinear partial differential equation in the gravitational potentials. In Sect. \ref{sect3}. we study the symmetry structure of the boundary equation and present the symmetry generators. Using a  general linear combination of the symmetry generators, we reduce the boundary condition to an ordinary differential equation. In  Sect. \ref{sect4}. we present several new families of exact solutions. In Sect. \ref{sect5}. we regain previous models as special cases and show that the new potentials derived satisfy a linear equation of state for certain parametric restrictions. In Sect. \ref{sect6}. we conclude by briefly discussing the findings of the analysis, and comment on the new solutions obtained.

\section{Basic equations}\label{sect2}
The collapsing interior of a radiating star satisfies the metric 
\begin{equation}\label{1}
	ds^{2}=-A^{2}dt^{2}+B^{2}dr^{2}+R^{2}(d\theta^{2}+\sin^{2}\theta d\phi^{2}),
\end{equation} 
where $A, B$ and $R$ are functions of the coordinate radius $r$ and  time $t$, representing gravity. The acceleration vector, expansion scalar and the shear scalar for the metric \eqref{1} are given by 
\begin{equation}\label{kinematics}
\dot{u}^{a}=\left(0,\frac{A^{\prime}}{AB^2},0,0\right),\hspace{0.1cm}\Theta=\frac{1}{A}\left(\frac{\dot{B}}{B}+2\frac{\dot{R}}{R}\right),\hspace{0.1cm}\sigma=-\frac{1}{3A}\left(\frac{\dot{B}}{B}-\frac{\dot{R}}{R}\right),
\end{equation}
respectively. Dots and primes represent differentiation with respect to time and the radial coordinate, respectively.
 The matter distribution is described by the energy momentum tensor which has the form
\begin{equation}
	T_{ab}=(\mu+p)u_{a}u_{b}+pg_{ab}+q_{a}u_{b}+q_{b}u_{a}+\pi_{ab},
\end{equation} 
where $u^{a}=(\frac{1}{A},0,0,0)$ denotes the fluid velocity, $q^{a}=(0,Bq,0,0)$ denotes the heat flux, $p$ is the isotropic pressure and $\pi^{ab}$ represents the stress tensor. The isotropic pressure has the form 
\begin{equation}
	p=\frac{1}{3}(p_{\parallel}+2p_{\perp}),
\end{equation}
where $p_{\parallel}$ and $p_{\perp}$ represent the radial and tangential pressures, respectively. For these particular forms of the energy momentum tensor, fluid velocity, heat flux and pressure, the Einstein field equations, for the metric \eqref{1}, become
\begin{subequations}\label{2}
	\begin{eqnarray}
		\mu&=&\frac{2}{A^{2}}\frac{\dot{B}}{B}\frac{\dot{R}}{R}+\frac{1}{R^{2}} +\frac{1}{A^{2}}\frac{\dot{R}^{2}}{R^{2}}-\frac{1}{B^{2}}\left(2 \frac{R^{\prime\prime}}{R}+\frac{R^{\prime 2}}{R^{2}}-2\frac{B^{\prime}}{B}\frac{R^{\prime}}{R}\right), \label{2.1}  \\
		p_ \parallel&=&\frac{1}{A^2}\left( -2\frac{\ddot{R}}{R}-\frac{\dot{R}^{2}}{R^{2}}+2\frac{\dot{A}}{A}\frac{\dot{R}}{R}\right)+\frac{1}{B^{2}}\left( \frac{R^{\prime 2}}{R^{2}}+2\frac{A^{\prime}}{A}\frac{R^{\prime}}{R}\right)-\frac{1}{R^{2}},\label{2.2} \\
		p_ \perp&=&-\frac{1}{A^{2}}\left( \frac{\ddot{B}}{B}-\frac{\dot{A}}{A}\frac{\dot{B}}{B}+\frac{\dot{B}}{B}\frac{\dot{R}}{R}-\frac{\dot{A}}{A}\frac{\dot{R}}{R}+\frac{\ddot{R}}{R}\right)\nonumber\\
		&&+\frac{1}{B^2}\left( \frac{A^{\prime\prime}}{A}-\frac{A^{\prime}}{A}\frac{B^{\prime}}{B}+\frac{A^{\prime}}{A}\frac{R^{\prime}}{R}-\frac{B^{\prime}}{B}\frac{R^{\prime}}{R}+\frac{R^{\prime\prime}}{R}\right),  \label{2.3}\\
		q&=&-\frac{2}{AB}\left(-\frac{\dot{R}^{\prime}}{R} +\frac{\dot{B}}{B}\frac{R^{\prime}}{R}+\frac{A^{\prime}}{A}\frac{\dot{R}}{R}\right),\label{2.4} 
	\end{eqnarray}
\end{subequations}

The exterior of the star is governed by the metric
\begin{equation}\label{5}
	ds^{2}=-\left(1-\frac{2m(v)}{\mathcal{R}} \right)dv^{2}-2dvd\mathcal{R}+\mathcal{R}^{2}\left( d\theta^{2}+\sin^{2}\theta d\phi^{2} \right), 
\end{equation}
which is the famous Vaidya solution. The matching of the potentials and the extrinsic curvature at the stellar surface generates the junction conditions. The relevant junction condition for the subsequent analysis is
\begin{equation}\label{bcon}
	\left(p_{\parallel}\right)_{\Sigma}=\left(q\right)_{\Sigma},
\end{equation}
where $\Sigma$ represents the boundary of the star. Substituting \eqref{2.2} and \eqref{2.4} into \eqref{bcon} implies that the potentials, on $\Sigma$, must satisfy 
\begin{eqnarray} \label{mastereqn} 
	&&2AB^{2}R\ddot{R}+AB^{2}\dot{R}^{2}-2B^{2}R\dot{A}\dot{R}-2ABRA^{\prime}\dot{R}+2A^{2}BR\dot{R}^{\prime}-2A^{2}RA^{\prime}R^{\prime}\nonumber\\
	&&-2A^{2}R\dot{B}R^{\prime}-A^{3}R^{\prime 2}+A^{3}B^{2}=0.
\end{eqnarray}
This is the master equation describing the general relativistic evolution of a star that is accelerating, expanding and shearing. Equation \eqref{mastereqn} is a highly nonlinear equation but solutions to it do exist; the first exact model was given by Thirukkanesh et al. \cite{5}.

 \section{Analysis} \label{sect3} 
 The master equation  is an under determined nonlinear partial differential equation and due to its complexity, it cannot be solved by standard techniques. Therefore, we undertake a group analysis to seek solutions to \eqref{mastereqn}. We follow the approach and notation of Abebe et al. \cite{16} to generate the Lie symmetry generators that leave equation \eqref{mastereqn} invariant. The determination of the generators in general is described by Head \cite{head} with the related software package PROGRAM LIE. Using PROGRAM LIE, we find that \eqref{mastereqn} admits three symmetry generators represented by the vector fields
 
 \begin{subequations}\label{10}
 	\begin{eqnarray}
 		X_{1}&=&A\dot{\beta}(t)\frac{\partial}{\partial A}-\beta(t)\frac{\partial}{\partial t},\label{10.1}\\
 		X_{2}&=&B\alpha^{\prime}(r)\frac{\partial}{\partial B}-\alpha(r) \frac{\partial}{\partial r},\label{10.2}\\
 		X_{3}&=&A\frac{\partial}{\partial A}+B\frac{\partial}{\partial B}+R\frac{\partial}{\partial R},\label{10.3}
 	\end{eqnarray}
 \end{subequations}
 where $\beta(t)$ and $\alpha(r)$ are arbitrary functions of their arguments. 
 
 For the current analysis of \eqref{mastereqn} we utilize the symmetry generated by
 \begin{equation}\label{11}
 	X=a_{1}X_{1}+a_{2}X_{2}+a_{3}X_{3},
 \end{equation}
 where $a_{1},a_{2}\neq 0$ and $a_{3}$ are real constants. It should be noted that \eqref{11} is a generalization of the symmetry vector used by Abebe et al. \cite{16} to obtain solutions to \eqref{mastereqn}.  The differential invariants of \eqref{11} constitute the reduction variables in the reduction procedure; they can be constructed by solving the associated Lagrange's system
 \begin{equation}\label{13}
 	\frac{dt}{-a_{1}\beta(t)}=\frac{dr}{-a_{2}\alpha(r)}=\frac{dA}{A(a_{1}\dot{\beta}(t)+a_{3})}=\frac{dB}{B(a_{2}\alpha^{\prime}(r)+a_{3})}=\frac{dR}{a_{3}R}.
 \end{equation}
 This is a system of ordinary differential equations. We find that the new independent variable is 
 \begin{equation}\label{15}
 	x=\frac{1}{a_{2}}\int\frac{dr}{\alpha(r)}-\frac{1}{a_{1}}\int\frac{dt}{\beta(t)}.
 \end{equation}  
 The gravitational potentials become
 \begin{subequations}\label{16}
 	\begin{eqnarray}
 		A&=&f(x)F_{1}(t),\\
 		B&=&g(x)F_{2}(r),\\
 		R&=& h(x)e^{-\frac{a_{3}}{a_{2}}\int\frac{dr}{\alpha(r)}},
 	\end{eqnarray}
 \end{subequations}
 where $f(x),h(x)$ and $g(x)$ are arbitrary functions of $x$. We have defined
 \begin{subequations}\label{functions}
 	\begin{eqnarray}
 		F_{1}(t)&=&\frac{e^{-\frac{a_{3}}{a_{1}}\int\frac{dt}{\beta(t)}}}{\beta(t)},\\
 		F_{2}(r)&=&\frac{e^{-\frac{a_{3}}{a_{2}}\int\frac{dr}{\alpha(r)}}}{\alpha(r)},
 	\end{eqnarray}
 \end{subequations}
 as new functions.
 
 The transformed quantities \eqref{15}--\eqref{functions} are new and do not appear in any previous analysis of a radiating star or analysis of the basic equation \eqref{mastereqn}. Therefore the new variables will lead to new solutions of the boundary condition \eqref{mastereqn}. The appearance of the exponential terms in \eqref{16} and \eqref{functions} increases the complexity and nonlinearity of the model but does allow for more general behaviour. If we set $a_{1}=-a_{4}$, where $a_{4}\neq0$ is some real number, $a_{2}=1$ and $a_{3}=0$, then we get  
 
 \begin{subequations}\label{abebecase}
 	\begin{eqnarray}
 		x&=&\int\frac{dr}{\alpha(r)}+\frac{1}{a_4}\int\frac{dt}{\beta(t)},\\
 		A&=&\frac{f(x)}{\beta(t)},\\
 		B&=&\frac{g(x)}{\alpha(r)},\\
 		R&=& h(x),
 	\end{eqnarray}
 \end{subequations}
 which regains the special case of Abebe et al. \cite{16} for a generalized Euclidean star. The analysis of \eqref{15}--\eqref{functions} should therefore lead to new classes of solutions to the boundary condition for a radiating star.
 
 Substituting for $A,B$ and $R$ from \eqref{16}, as well as their respective derivatives, into the master equation \eqref{mastereqn} leads to
 \begin{align}\label{riccati}
 	g^{\prime}(x)+W(x)g(x)^2+Q(x)g(x)+S(x) =0,
 \end{align} 
 where the coefficient functions are 
 \begin{subequations}\label{17}
 	\begin{eqnarray}
 		W(x)&=&\frac{a_2e^{-a_3x}\left(a_1^2e^{2a_3x}f^3+f\left(2h\left( h''-a_3h'\right)+h'^2\right)-2f'hh'\right)}{2a_1f^2h\left(h'-a_3h\right)},\label{17.1}\\
 		Q(x)&=&\frac{f h''-h'\left(a_3 f+f'\right)}{f \left(a_3 h-h'\right)},\label{17.2}\\
 		S(x)&=&-\frac{a_1e^{a_3x} \left(f\left(h'-a_3 h\right)+2f^{\prime}h\right)}{2a_2 h}.\label{17.3}
 	\end{eqnarray}
 \end{subequations}
 Note that primes in equations \eqref{riccati} and \eqref{17} now imply differentiation with respect to the variable $x$. We have written \eqref{riccati} as a Riccati equation in the function $g(x)$. It is also possible to write \eqref{riccati} as a function of $f(x)$ or $h(x)$ but the resulting ordinary differential equations are not easily solvable. The form \eqref{riccati} proves to be the most useful. 
 
  \section{New solutions}\label{sect4}
  The reduced master equation \eqref{riccati} cannot be solved in general. However, by placing certain conditions on the coefficient functions in \eqref{17} will result in simplification and possibly lead to exact solutions to \eqref{riccati}. To this end we investigate restrictions on the coefficient functions separately and solve \eqref{riccati} in four cases.
  
  \subsection{Case $1$: $S(x)=0$}
  In this case \eqref{riccati} becomes a homogeneous equation of Bernoulli form. The condition $S(x)=0$ implies that $f(x)$ and  $h(x)$ must satisfy the differential equation
  \begin{equation}\label{relation1}
  	h^{\prime}+\left(2\frac{f^{\prime}}{f}-a_3\right)h=0.
  \end{equation} 
  This equation can be solved for $h$, in terms of $f$, to obtain
  \begin{equation}\label{relation1}
  	h(x)=\frac{C_{1}e^{a_{3}x}}{f(x)^2}.
  \end{equation}
  Using the relationship \eqref{relation1} in equation \eqref{riccati} yields
  \begin{eqnarray}\label{19}
  	&&g^{\prime}-a_2\left(\frac{a_3^2 C_1^2f^2+a_1^2f^8+20C_1^2f^{\prime 2}-2C_1^2f(2f^{\prime\prime}+5a_3f^{\prime})}{4a_{1}C_1^2e^{a_3x} f^2 f'}\right)g^2\nonumber\\
  	&&-\left(\frac{f^{\prime\prime}}{f^{\prime}}-4\frac{f^{\prime}}{f}+\frac{3a_3}{2}\right)g=0.
  \end{eqnarray}
  The simplified equation \eqref{19} can be solved for $g(x)$, and we find that 
  \begin{eqnarray}\label{20}
  	g(x)&=& \frac{e^{\frac{3a_3}{2}x}f^{\prime}(x)}{f(x)^{4}}\bigg[C_2-a_2\int^xe^{\frac{a_3}{2}w}\bigg(\frac{(a_3C_1f(w))^2+a_1^2f(w)^8}{4a_1C_1^2f(x)^6}\nonumber\\
  	&&-\frac{10a_3C_1^2f(w)f^{\prime}(w)+20C_1^2f^{\prime}(w)^2-4C_1^2f(w)f^{\prime\prime}(w)}{4a_1C_1^2f(x)^6}\bigg)dw\bigg]^{-1},
  \end{eqnarray}
  where $C_{1}$ and $C_{2}$ are constants of integration. We have obtained the general solution to \eqref{riccati} when $S(x)=0$.
  
  Upon substituting \eqref{relation1} and relation \eqref{20} into \eqref{16}, we find that the gravitational potentials \eqref{16} can be written as
  \begin{subequations}\label{pot1}
  	\begin{eqnarray}
  		A&=& f(x)F_{1}(t),\\
  		B&=&  \frac{e^{\frac{3a_3}{2}x}f^{\prime}(x)F_{2}(r)}{f(x)^{4}}\bigg[C_2-a_2\int^xe^{\frac{a_3}{2}w}\bigg(\frac{(a_3C_1f(w))^2+a_1^2f(w)^8}{4a_1C_1^2f(x)^6}\nonumber\\
  		&&-\frac{10a_3C_1^2f(w)f^{\prime}(w)+20C_1^2f^{\prime}(w)^2-4C_1^2f(w)f^{\prime\prime}(w)}{4a_1C_1^2f(x)^6}\bigg)dw\bigg]^{-1},\\
  		R&=&\frac{C_{1}e^{a_{3}\left(x-\frac{1}{a_{2}}\int\frac{dr}{\alpha(r)}\right)}}{f(x)^2}.
  	\end{eqnarray}  
  \end{subequations}
  The gravitational potentials in \eqref{pot1} are in terms of an arbitrary function $f(x)$ and constitute a new family of solutions. 
  
  In this case the heat flux becomes
  \begin{equation}
  q=-\frac{4 e^{-\frac{3a_3}{2}x}f(x)^2 H(x)}{a_1\alpha (r)F_1(t) \beta (t) F_2(r)},
  \end{equation}
  where we have set 
  \begin{eqnarray}
  H(x)&=&\frac{(a_3C_1f(x))^2+a_1^2f(x)^8-10a_3C_1^2f(x)f^{\prime}(x)}{4a_1C_1^2f(x)^6}\nonumber\\
  &&+\frac{4C_1^2f(x)f^{\prime\prime}(x)-20C_1^2f^{\prime}(x)^2}{4a_1C_1^2f(x)^6}.
  \end{eqnarray}

  \subsection{Case $2$: $W(x)=0$}
  
  In this case \eqref{riccati} is a linear equation. Imposing the condition $W(x)=0$ implies the relationship
  \begin{equation}\label{frel}
  	a_1^2e^{2a_3x}f^3+f\left(h\left(2 h''-2a_3h'\right)+h'^2\right)-2hf'h'=0.
  \end{equation}
  We can solve \eqref{frel} for $f(x)$ and find that the solution is 
  \begin{equation}\label{33}
  	f(x)=\frac{e^{-a_{3}x}\sqrt{h(x)} h'(x)}{\sqrt{C_1-a_{1}^2 h(x)}}.
  \end{equation}
  Substituting \eqref{33} into \eqref{riccati} produces a  first order linear equation of the form
  \begin{eqnarray}\label{34}
  	&&g^{\prime}-a_1\left(\frac{h'\left(3a_{3}h \left(a_{1}^2h-C_1\right)+h'\left(2 C_1-a_{1}^2 h\right)\right)+2hh''\left(C_1-a_{1}^2h\right)}{2a_{2}\sqrt{h} \left(C_1-a_{1}^2 h\right)^{\frac{3}{2}}}\right)\nonumber\\
  	&&+\left(\frac{C_1h^{\prime 2}}{2h(C_1-a_1^2h)(h^{\prime}-a_3h)}\right)g =0.
  \end{eqnarray}
  The solution to equation \eqref{34} is given by 
  \begin{eqnarray}\label{36}
  	g(x)&=&\exp\left(-\int^{x}\frac{C_1h'(w)^2}{2h(w)\left(C_1-a_{1}^2 h(w)\right)\left(h'(w)-a_{3} h(w)\right)}dw\right)\nonumber\\
  	&&\times\Bigg[C_{2}+\int^x\bigg[\exp\bigg(\int^{w}\frac{C_1h'(z)^2}{2h(z)\left(C_1-a_{1}^2 h(z)\right)\left(h'(z)-a_{3} h(z)\right)}dz\bigg)\nonumber\\
  	&&\bigg(\frac{a_{1}h'(w)\left(3a_{3}h(w)\left(a_{1}^2h(w)-C_1\right)+h'(w)\left(2 C_1-a_{1}^2h(w)\right)\right)}{2a_{2} \sqrt{h(w)}\left(C_1-a_{1}^2 h(w)\right)^{\frac{3}{2}}}\nonumber\\
  	&&+\frac{2 h(w) h''(w)\left(C_1-a_{1}^2 h(w)\right)}{2a_{2} \sqrt{h(w)} \left(C_1-a_{1}^2 h(w)\right)^{\frac{3}{2}}}\bigg)\bigg]dw\Bigg].
  \end{eqnarray}
  Hence we have obtained the general solution to \eqref{riccati} when $W(x)=0$.
  
  On substituting \eqref{33} and \eqref{36} into \eqref{16}, the gravitational potentials become
  \begin{subequations}\label{W}
  	\begin{eqnarray}
  		A&=&\frac{e^{-a_{3}x}\sqrt{h(x)} h'(x)F_{1}(t)}{\sqrt{C_1-a_{1}^2h(x)}},\\
  		B&=& F_{2}(r)\exp\left(-\int^{x}\frac{C_1h'(w)^2}{2h(w)\left(C_1-a_{1}^2 h(w)\right)\left(h'(w)-a_{3} h(w)\right)}dw\right)\nonumber\\
  		&&\times\Bigg[C_{2}+\int^x\bigg[\exp\bigg(\int^{w}\frac{C_1h'(z)^2}{2h(z)\left(C_1-a_{1}^2 h(z)\right)\left(h'(z)-a_{3} h(z)\right)}dz\bigg)\nonumber\\
  		&&\bigg(\frac{a_{1}h'(w)\left(3a_{3}h(w)\left(a_{1}^2h(w)-C_1\right)+h'(w)\left(2 C_1-a_{1}^2h(w)\right)\right)}{2a_{2} \sqrt{h(w)}\left(C_1-a_{1}^2 h(w)\right)^{\frac{3}{2}}}\nonumber\\
  		&&+\frac{2 h(w) h''(w)\left(C_1-a_{1}^2 h(w)\right)}{2a_{2} \sqrt{h(w)} \left(C_1-a_{1}^2 h(w)\right)^{\frac{3}{2}}}\bigg)\bigg]dw\Bigg],\\
  		R&=& h(x)e^{-\frac{a_{3}}{a_{2}}\int\frac{dr}{\alpha(r)}}.
  	\end{eqnarray}
  \end{subequations}
  The potentials in \eqref{W} are in terms of an arbitrary function $h(x)$ and generate a new family of exact solutions.
 
 For the potentials \eqref{W} the heat flux is
 \begin{eqnarray}
 q&=&\frac{e^{a_3 x+\int H(x) \, dx}}{a_1 a_2 F_2(r) F_1(t)  \alpha (r) \beta (t)h(x)^{\frac{5}{2}} h'(x) \sqrt{C_1-a_1^2 h(x)} \left(C_2+\int G(x) \, dx\right){}^2}\nonumber\\
 &&\times\Bigg[2h(x)\left(a_3h(x)-h'(x)\right) \left(a_1^2 h(x)-C_1\right)\bigg(G(x)-H(x) \bigg(C_2\nonumber\\
 &&+\int^{x} G(w)dw\bigg)\bigg)+c_1 h'(x)^2 \left(C_2+\int^{x} G(w)dw\right)\Bigg],
 \end{eqnarray} 
  where 
  \begin{equation}
  H(x)=\frac{C_1h'(x)^2}{2h(x)\left(C_1-a_{1}^2 h(x)\right)\left(h'(x)-a_{3} h(x)\right)},
  \end{equation}
   and 
   \begin{eqnarray}
  G(x)&=& \Bigg(\frac{a_{1}e^{a_{3}x}h'(x)\left(3a_{3}h(x)\left(a_{1}^2h(x)-C_1\right)+h'(x)\left(2 C_1-a_{1}^2h(x)\right)\right)}{2a_{2} \sqrt{h(x)}\left(C_1-a_{1}^2 h(x)\right)^{\frac{3}{2}}}\nonumber\\
   &&+\frac{2 h(x) h''(x)\left(C_1-a_{1}^2 h(x)\right)}{2a_{2} \sqrt{h(x)} \left(C_1-a_{1}^2 h(x)\right)^{\frac{3}{2}}}\Bigg)\exp\bigg(\int^{x}H(w)dw-a_{3}x\bigg).
   \end{eqnarray}

  \subsection{Case $3$: $Q(x)=0$}\label{4.3}
  In this case \eqref{riccati} is a Riccati equation. With the restriction $Q(x)=0$, we obtain the constraint that $f(x)$ and $h(x)$ must satisfy the differential equation
  \begin{equation}\label{relation2}
  	f^{\prime}+\left(a_3-\frac{h^{\prime\prime}}{h^{\prime}}\right)f=0.
  \end{equation}
  Solving \eqref{relation2}, we find that 
  \begin{equation}\label{relation3}
  	f(x)=C_1e^{-a_{3}x} h'(x).
  \end{equation}
  Substituting \eqref{relation3} into \eqref{riccati} results in the simpler equation
  \begin{equation}\label{specialriccati}
  	g^{\prime}-\left(\frac{a_{2}\left(a_{1}^2 C_1^2+1\right) h'}{2a_{1}C_1 h\left(a_{3} h-h'\right)}\right)g^2-\frac{a_{1}C_1 \left(h\left(2 h''-3a_{3}h'\right)+h'^2\right)}{2a_{2}h}=0,
  \end{equation}
  which is a Riccati equation that cannot be solved in general. However, it is possible to find special solutions if the function $h(x)$ is specified. We demonstrate this for a simple choice below.
  
  If we assume the form
  \begin{equation}
  	h(x)=ke^{x},
  \end{equation}
  then \eqref{relation3} becomes
  \begin{equation}\label{fsol}
  	f(x)=kC_1e^{(1-a_3)x},
  \end{equation}
  and we find that \eqref{specialriccati} simplifies to
  \begin{equation}\label{neweq}
  	g^{\prime}(x)=\left(\frac{a_2(a_1^2C_1^2+1)}{2a_1kC_1(a_3-1)}\right)e^{-x}g(x)^2-\left(\frac{3a_1kC_1(a_3-1)}{2a_2}\right)e^{x}.
  \end{equation}
  We can solve \eqref{neweq}  to find
  \begin{equation}\label{gsol}
  	g(x)=\frac{3a_1k C_1(a_3-1)e^{x}\left(1+\gamma (1+C_2(1-\gamma)e^{\gamma x})\right)}{a_2(\gamma^2-1)(1+\gamma C_2e^{\gamma x})},
  \end{equation}
  where $\gamma=\sqrt{4+3a_1^2C_1^2}$.
  
  In this case the potentials become
  \begin{subequations}\label{newpots}
  	\begin{eqnarray}
  		A&=& kC_1e^{(1-a_3)x}F_{1}(t),\\
  		B&=&\frac{3a_1k C_1(a_3-1)e^{x}\left(1+\gamma(1+C_2(1-\gamma)e^{\gamma x})\right)}{a_2(\gamma^2-1)(1+\gamma C_2e^{\gamma x})}F_{2}(r),\\
  		R&=& ke^{\left(x-\frac{a_{3}}{a_{2}}\int\frac{dr}{\alpha(r)}\right)}.
  	\end{eqnarray}
  \end{subequations}
  Thus we have found another family of exact solutions to the boundary condition \eqref{mastereqn}.  If we set $C_2=0$, then the shear scalar $\sigma=0$ from \eqref{kinematics} implying that the spacetime is not shearing. It should be noted that the shear-free solution \eqref{newpots} appears to be new. It is not contained in the family of solutions found by Abebe et al. \cite{shear-free} as in their model $a_{3}=0$ in \eqref{11} with only two infinitesimal generators. In our case $a_{3}\neq 0$ and there are three generators. The shear-free models of Herrera et al. \cite{22}, Maharaj and Govender \cite{23} and Herrera et al. \cite{24} are conformally flat. Our class of solution \eqref{newpots} does not satisfy this condition in general, and is therefore different. 
  
  For the gravitational potentials \eqref{newpots} we find that the heat flux is
  \begin{equation}
  q=\frac{2 \left(\gamma ^2-1\right) e^{\left(a_3-2\right) x} \left((\gamma -1) \gamma C_2  e^{\gamma  x} \left( \gamma C_2  e^{\gamma  x}+2 \gamma +2\right)-\gamma -1\right)}{3 a_1^2 C_1^2 k^2 F_1(t) \alpha (r) \beta (t) \left(\gamma+1-\gamma C_2 (\gamma -1)e^{\gamma  x}\right){}^2}.
  \end{equation}

  \subsection{Case 4: Linear relationship between $f(x)$ and $g(x)$}\label{4.4}
  In the previous cases we have solved \eqref{riccati} by analyzing the under determined ordinary differential equation in three cases. This was achieved by placing restrictions on the coefficients in \eqref{17} and generating functional forms for $g(x)$, $h(x)$ and $f(x)$. In the families of solutions found there were no direct relationships between the functions $f$, $g$ and $h$. The existence of a relationship may produce other exact solutions; the simplest connection is linear. In this treatment of \eqref{riccati}, we assume that $f(x)$ and $g(x)$ are linearly related by
  \begin{equation}\label{linearrelation}
  	f(x)=c g(x),
  \end{equation}
  where $c\neq0$ is a constant of proportionality. The relationship \eqref{linearrelation} transforms equation \eqref{riccati} into
  \begin{eqnarray}\label{39}
  	&& g^{\prime}-\left(\frac{a_{1}^2a_{2}^2c^2e^{2a_{3} x}}{2h\left(a_{2}-a_{1}ce^{a_{3}x}\right)\left(a_{1}c e^{a_{3}x}\left(a_{3}h-h'\right)+a_{2}h'\right)}\right) g^3\nonumber\\
  	&&+\frac{1}{2} \left(\frac{a_{2}\left(a_{3}h-2h'\right)}{h\left(a_{2}-a_{1}ce^{a_{3}x}\right)}+\frac{2a_{2}\left(a_{3} h'-h''\right)}{a_{1}ce^{a_{3}x}\left(a_3h-h'\right)+a_{2}h'}+\frac{h'}{h}-a_{3}\right) g=0.
  \end{eqnarray}
  Observe that \eqref{39} is a Bernoulli equation in $g(x)$. We can solve \eqref{39} to obtain the general solution
  \begin{eqnarray}\label{40}
  	g(x)&=&\frac{e^{\left(\frac{1}{2}\int^{x}H(w)dw\right)}}{\left(C_{1}-\int^{x}\frac{(2a_{1}a_{2}c)^{2}\exp\big(\int^{w} H(z)dz+2a_{3}w\big)}{2 h(w)(a_{2}-a_{1}ce^{a_{3}w})(a_{1}ce^{a_{3}w} \left(a_{3}h(w)-h'(w)\right)+a_{2} h'(w))}dw\right)^{\frac{1}{2}}},
  \end{eqnarray}
  where we have introduced the function
  \begin{eqnarray}
  	H(x)=&&\frac{a_{2}\left(a_{3}h(x)-2h'(x)\right)}{h(x)\left(a_{1}c e^{(a_{3}x)}-a_{2}\right)}-\frac{h'(x)}{h(x)}+a_{3}\nonumber\\
  	&&+\frac{2a_{2} \left(h''(x)-a_{3} h'(x)\right)}{a_{1}ce^{(a_{3}x)} \left(a_{3} h(x)-h'(x)\right)+a_{2} h'(x)}.
  \end{eqnarray}
  Thus we have produced another new family of exact solutions when $f(x)=cg(x)$.
  
  The gravitational potentials can be found explicitly. With the help of \eqref{40}, \eqref{linearrelation} and \eqref{16} we obtain
  \begin{subequations}\label{41}
  	\begin{eqnarray}
  		A&=& cB\frac{F_{1}(t)}{F_{2}(r)}\label{4},\\
  		B&=&\frac{F_{2}(r)e^{\left(\frac{1}{2}\int^{x}H(w)dw\right)}}{\left(C_{1}-\int^{x}\frac{(2a_{1}a_{2}c)^{2}\exp\big(\int^{w} H(z)dz+2a_{3}w\big)}{2 h(w)(a_{2}-a_{1}ce^{a_{3}w})(a_{1}ce^{a_{3}w} \left(a_{3}h(w)-h'(w)\right)+a_{2} h'(w))}dw\right)^{\frac{1}{2}}}\label{5},\\
  		R &=& \alpha(r)h(x)F_{2}(r).\label{6}
  	\end{eqnarray}
  \end{subequations}
  
  For this category of solution the heat flux becomes
  \begin{eqnarray}
  q&=&\Bigg[ G(x) \bigg(F_2(r) \left(a_2 h(x) \alpha '(r)+2 h'(x)\right)+a_2 h(x) \alpha (r) F_2'(r)\bigg)\nonumber\\
  &&-\left(\int^{x} G(w)dw-C_1\right) \bigg[H(x) \Big(F_2(r) \left(a_2 h(x) \alpha '(r)+2 h'(x)\right)\nonumber\\
  &&+a_2 h(x) \alpha (r) F_2'(r)\Big)-2 a_2 h'(x) \left(F_2(r) \alpha '(r)+\alpha (r) F_2'(r)\right)\nonumber\\
  &&-2 F_2(r) h''(x)\bigg]\Bigg]\frac{e^{-\int^x H(w) \, dw}}{a_1 a_2 c F_2(r){}^2 F_1(t) h(x) \alpha (r) \beta (t)},
  \end{eqnarray}
 where
 \begin{eqnarray}
 G(x)&=&\frac{(2a_{1}a_{2}c)^{2}\exp\big(\int^{x} H(w)dw+2a_{3}x\big)}{2 h(x)(a_{2}-a_{1}ce^{a_{3}x})(a_{1}ce^{a_{3}x} \left(a_{3}h(x)-h'(x)\right)+a_{2} h'(x))}.
 \end{eqnarray}
\section{Previous results and equations of state}\label{sect5}

The generalized Lie point symmetry \eqref{11} has led to several new solutions of the stellar boundary condition \eqref{mastereqn}. We can recover earlier results from our analysis. If we impose the conditions
\begin{equation}
	a_{1}=-a_{4},\;a_{2}=1,\;a_{3}=0,
\end{equation}
then the potentials \eqref{41} in Sect. \ref{4.4}. are substantially simplified. We obtain the reduced potentials
\begin{subequations}\label{42}
	\begin{eqnarray}
		A&=&\frac{ch^{\frac{1-ca_{4}}{2(ca_{4}+1)}}h^{\prime\frac{1}{ca_{4}+1}})}{\left(C_{1}-\left(\frac{ca_{4}}{ca_{4}+1}\right)^{2}\int^{x}h(w)^{-\frac{2ca_4}{ca_{4}+1}}h^{\prime}(w)^{\frac{1-ca_{4}}{ca_{4}+1}}\text{d}w\right)^{\frac{1}{2}}\beta(t)}\label{42.1},\\
		B&=&\frac{h^{\frac{1-ca_{4}}{2(ca_{4}+1)}}h^{\prime\frac{1}{ca_{4}+1}}}{\left(C_{1}-\left(\frac{ca_{4}}{ca_{4}+1}\right)^{2}\int^{x}h(w)^{-\frac{2ca_4}{ca_{4}+1}}h^{\prime}(w)^{\frac{1-ca_{4}}{ca_{4}+1}}\text{d}w\right)^{\frac{1}{2}}\alpha(r)}\label{42.2},\\
		R&=&h(x)\label{42.3}.
	\end{eqnarray}
\end{subequations}
The solution \eqref{42} corresponds to the generalized Euclidean star model of Abebe et al. \cite{16}. If we make the further restriction $B=\dfrac{\partial R}{\partial r}$, then we regain the Euclidean star formulation of Herrera and Santos \cite{14}. Furthermore, if we set $h(x)=mx^{n}$, then we can regain the potentials 
\begin{subequations}\label{euclidpots}
	\begin{eqnarray}
		A&=&\frac{x^{n-1}}{\beta(t)},\\
		B&=&\tilde{m}n\frac{x^{n-1}}{\alpha(r)},\\
		R&=&\tilde{m}x^{n},
	\end{eqnarray}
\end{subequations}
where $x=\int\frac{dr}{\alpha(r)}+\frac{1}{a_4}\int\frac{dt}{\beta(t)}$, $\tilde{m}=m^{n}$ and $\tilde{m}^{2}n^{3}-2\tilde{m}n^{2}+2n(\tilde{m}-1)+2=0$. This exact model for a general relativistic Euclidean star  was found by Govinder and Govender \cite{15}.

Particular models in our general class of solutions admit a barotropic equation of state. We point out two examples in the absence of shear and in the presence of shear. 	As a first example consider the model generated in Sect. \ref{4.3}.. If we set $C_{2}=0$ in the potentials \eqref{newpots}, then $\dot{u}^{a}\neq0,\;\Theta\neq0$ and $\sigma=0$: the spacetime is expanding and accelerating but shear-free. The energy density and radial pressure are given by
\begin{subequations}
	\begin{eqnarray}
		\mu&=&\frac{2(\gamma^2+\gamma+7)e^{\left(\frac{2a_{3}}{a_{2}}\int\frac{dr}{\alpha(r)}-2x\right)}}{9a_1^2k^2C_1^2}\label{dens},\\
		p_{\parallel}&=&-\frac{2(\gamma-1)e^{\left(\frac{2a_{3}}{a_{2}}\int\frac{dr}{\alpha(r)}-2x\right)}}{3a_1^2k^2C_1^2}\label{heat}.
	\end{eqnarray}
\end{subequations}
These quantities satisfy the relation
\begin{equation}
	p_{\parallel}=3\left(\frac{1-\gamma}{\gamma^2+\gamma+7}\right)\mu,\;\gamma=\sqrt{4+3a_1^2C_1^2},
\end{equation}
which is a linear equation of state. Hence this is an example of a radiating star which is shear-free with a barotropic equation of state. Note that this example is not contained in the shear-free models of Abebe et al. \cite{shear-free} as the infinitesimal Lie generator \eqref{11} is different in our case and $a_3\neq 0$. As a second example consider the model generated in Sect. \ref{4.4}. which contains the generalized Euclidean model \eqref{42}. For this example $\dot{u}^{a}\neq0,\;\Theta\neq0$ and $\sigma\neq0$: the spacetime is expanding, accelerating and shearing. The energy denisty and radial pressure have the form 
\begin{subequations}
	\begin{eqnarray}
		\mu&=& \frac{ 2\left(1-a_4^3 c^3\right)\varphi (x) h^{\frac{a_4c-1}{1+a_4c}}\left(h'^2+h h''\right)}{ a_4^2 c^2(1+a_4 c)^3 h^2h'^{\frac{2}{1+a_4 c}}}\nonumber \\
		&&+\frac{2a_4^2 c^2(1+a_4 c) (1+a_4 c (1+a_4 c))h'^{\frac{2}{1+a_4 c}}}{a_4^2 c^2(1+a_4 c)^3 h^2h'^{\frac{2}{1+a_4 c}}},\\
		p_\parallel&=& \frac{2 \left((a_4 c-1)\varphi (x) h^{\frac{a_4c-1}{1+a_4 c}}\left(h'^2+h h''\right)-a_4^2 c^2 (1+a_4 c)h'^{\frac{2}{1+a_4 c}}\right)}{a_4 c (1+a_4 c)^3 h^2h'^{\frac{2}{1+a_4 c}}},
	\end{eqnarray}
\end{subequations}
where $\varphi (x)=(1+a_4c)^2 d-(a_4c)^2 \int^{x} h(w)^{-\frac{2 a_4c}{1+a_4c}} h'(w)^{\frac{1-a_4c}{1+a_4c}} dw$. These quantities obey the relation
\begin{equation}
	p_\parallel(\mu)=\lambda \mu,\; \lambda =-\frac{a_4c}{(a_4c)^2 +a_4c +1},
\end{equation}
which is also linear. Hence this is an example of a radiating star which is shearing with a barotropic matter distribution. This equation of state was also identified by Abebe et al. \cite{16}.

The energy conditions in the presence of anisotropic pressures and heat flux was developed by Chan \cite{10}. For special choices of the Euclidean metric  \eqref{euclidpots} and the generalized Euclidean potentials \eqref{42} we can show that $E_{\text{wec}}=\mu- p_{\parallel} + \sqrt{(\mu+ p_{\parallel} )^2 - 4 q^2}\geq0$, $E^{(1)}_{\text{dec}}=\mu - p_{\parallel}\geq0$, $E^{(2)}_{\text{dec}}= \mu - p_{\parallel} - 2p_{\perp} + \sqrt{(\mu+ p_{\parallel} )^2 - 4 q^2}\geq0$ and $ E_{\text{sec}}=2p_{\perp} + \sqrt{(\mu+ p_{\parallel} )^2 - 4 q^2}\geq0$ are valid. Therefore the weak, dominant and strong energy conditions are satisfied. This indicates that matter distribution in our new class of radiating stars is physically acceptable.

To study the dissipative collapse of a radiating star it is necessary to investigate the causal transport equations. Martinez \cite{25} and Di Prisco et al. \cite{26} discuss the importance of the transport equation in describing the gravitational evolution of a dissipating relativistic star. The Maxwell-Cattaneo heat transport equation for the line element \eqref{1} is 
\begin{equation}\label{transport}
\tau h_{a}^{\;b}\dot{q}_{b}+q_{a}=-\kappa\left(h_{a}^{\;b}\nabla_{b}T+T\dot{u}_{a}\right),
\end{equation}
where $\kappa$ is the thermal conductivity, $h_{ab}=g_{ab}+u_{a}u_{b}$ is the projection tensor, $\tau$ is the relaxation time and $T$ is the local temperature. Equation \eqref{transport} becomes the usual Fourier heat transport equation when $\tau=0$. It can be written as 
\begin{equation}\label{temp}
T(r,t)=-\frac{1}{\kappa A}\int\left[\tau(qB)^{\dot{}} B+AqB^{2}\right]dr.
\end{equation}
For each of the solutions presented in this paper we have explicit analytic forms for the heat flux $q$ and the potentials $A$ and $B$. Therefore it is possible to find the causal temperature $T$ from \eqref{temp} and investigate the role of the relaxation time $\tau$ on the thermal evolution in gravitational collapse. This is demonstrated in the analysis of Thirukkanesh et al. \cite{27} for example.
 \section{Discussion}\label{sect6}
 We have studied the boundary condition at the stellar surface for a radiating star in general relativity. We have extended previous analyses using the Lie symmetry approach by taking a general combination of all symmetry vectors admitted by the junction condition \eqref{mastereqn}. We find that the metric potentials all depend on exponential functions which increase the complexity of the model. This property with exponential functions is not present in other treatments with Lie symmetries in radiating stars. The boundary condition can the be written as a Riccati equation. Four new families of exact solutions to the Riccati equation are identified. We regain Euclidean and generalized Euclidean stellar models when the parameter $a_3$ vanishes. We show that linear equations of state are admitted for spacetimes which are shear-free and shearing. As our families of solutions contain the Govinder and Govender \cite{15} and Abebe et al. \cite{16} models, the energy conditions are satisfied which is a desirable feature in a physical model.

\begin{acknowledgements}
RM, AKT and RN thank the  University of KwaZulu-Natal and the National Research Foundation  for financial support.
SDM acknowledges that this work is based upon research supported by the South African Research Chair Initiative of the Department of Science and Technology and the National Research Foundation.
\end{acknowledgements}


\begin{thebibliography}{}



	
	\bibitem{1} Kolassis,  C. A., Santos,  N. O.,  Tsoubelis,  D.: Astrophys. J. \textbf{327}, 755 (1988)
	\bibitem{2}  Sharma, R., Das, S.: J. Grav. \textbf{2013}, 659605 (2013)
	\bibitem{3}  Sarwe, S., Tikekar, R.: Int. J. Mod. Phys. D \textbf{19}, 1889 (2010)
	\bibitem{4} Reddy,  K. P., Govender, M., Maharaj,  S. D.: Gen. Relativ. Gravit. \textbf{47}, 35 (2015)
	\bibitem{5}   Thirukkanesh, S., Rajah, S. S., Maharaj,  S. D.: J. Math. Phys. \textbf{53}, 032506 (2012)
	\bibitem{6}Govender, G., Govender, M., Govinder, K. S.: Int. J. Mod. Phys. D \textbf{19}, 1773 (2010)
	
	\bibitem{7}Govender,  M., Govinder, K. S., Fleming,  D.: Int. J. Theor. Phys. \textbf{51}, 3399 (2012)
	
	\bibitem{8} Herrera,  L., Di Prisco, A., Ospino, J.: Gen. Relativ. Gravit. \textbf{42}, 1585 (2010) 
	\bibitem{9}  Chan, R.: Mon. Not. R. Astron. Soc. \textbf{316}, 588 (2000)
	\bibitem{10} Chan,  R.: Astron. Astrophys. \textbf{368}, 325 (2001)
	\bibitem{11} Chan, R.:  Int. J. Mod. Phys. D \textbf{12}, 1131 (2003)
	\bibitem{12}Pinheiro,  G., Chan,  R.: Gen. Relativ. Gravit. \textbf{40}, 2149 (2008)
	\bibitem{13} Pinheiro,  G., Chan,   R.: Int. J. Mod. Phys. D \textbf{19}, 1797 (2010)
	\bibitem{14} Herrera, L., Santos,  N. O.: Gen. Relativ. Gravit. \textbf{42}, 2383 (2010)
	\bibitem{15}  Govinder,  K. S., Govender,  M.: Gen. Relativ. Gravit. \textbf{44}, 147 (2012)
	\bibitem{16} Abebe,  G. Z.,  Maharaj, S. D., Govinder, K. S.: Gen. Relativ. Gravit. \textbf{46}, 1733 (2014)
	\bibitem{17} Ivanov,  B. V.: Int. J. Mod. Phys. D \textbf{25}, 1650049 (2016)
	\bibitem{quant}Gagnon,  L., Winternitz,  P.: J. Phys. A: Math. Gen. \textbf{22}, 469 (1989)
	\bibitem{msomi} Msomi, A. M.,  Govinder,  K. S., Maharaj,  S. D.: J. Phys. A: Math. Theor. \textbf{43}, 285203 (2010)
	\bibitem{head}  Head, A. K.: Comput. Phys. Commun. \textbf{71}, 241 (1993)
	\bibitem{shear-free}  Abebe, G. Z., Maharaj,  S. D.,  Govinder, K. S.: Eur. Phys. J. C  \textbf{75}, 496 (2015)
	\bibitem{22} Herrera, L., Le Denmat, G., Santos, N. O., Wang, A.: Int. J. Mod. Phys. D \textbf{13}, 583 (2004)
	\bibitem{23} Maharaj, S. D., Govender, M.: Int. J. Mod. Phys. D \textbf{14}, 667 (2005)
	\bibitem{24} Herrera, L., Di Prisco, A., Ospino, J.: Phys. Rev. D \textbf{74}, 044001 (2006)
	\bibitem{25} Martinez, J.: Phys. Rev. D \textbf{53}, 6921 (1996)
	\bibitem{26} Di Prisco, A., Herrera, L., Esculpi, M.: Class. Quantum Grav. \textbf{13}, 1053 (1996)
	\bibitem{27} Thirukkanesh, S., Rajah, S. S., Maharaj, S. D.: J. Math. Phys. \textbf{53}, 032506 (2012)



\end{thebibliography}
\end{document}